\documentclass{ieeetj}

\usepackage{amsmath,amssymb,amsfonts}
\usepackage{mathtools}
\usepackage{bm}
\usepackage{stmaryrd}

\usepackage{graphicx,color}
\usepackage{xcolor}

\usepackage{hyperref}
\hypersetup{hidelinks=true}

\usepackage{footmisc}
\usepackage{setspace}

\usepackage{placeins}

\usepackage{tabularx}
\usepackage{booktabs}

\usepackage{cleveref}
\Crefname{figure}{Fig.}{Figs.}

\usepackage{jabbrv}
\usepackage[sort&compress,square,numbers]{natbib}

\usepackage{siunitx}
\AtBeginDocument{\definecolor{tmlcncolor}{cmyk}{0.93,0.59,0.15,0.02}\definecolor{NavyBlue}{RGB}{0,86,125}}

\def\seclogo{}
\def\authorrefmark#1{\ensuremath{^{\textbf{#1}}}}

\jvol{TBA}
\pubyear{2024}

\DefineJournalPartialAbbreviation{Transactions}{Trans}

\DefineJournalException{IEEE Transactions on Audio Speech and Language Processing}{IEEE Trans. ASLP}

\DefineJournalException{IEEE/ACM Transactions on Audio, Speech, and Language Processing}{IEEE/ACM Trans. ASLP}

\DefineJournalException{IEEE/ACM Transactions on Audio Speech and Language Processing}{IEEE/ACM Trans. ASLP}

\DefineJournalException{Proceedings of the IEEE International Conference on Acoustics, Speech, and Signal Processing}{Proc. ICASSP}

\DefineJournalException{Proceedings of the IEEE International Conference on Acoustics, Speech and Signal Processing}{Proc. ICASSP}

\DefineJournalException{Proceedings of the IEEE International Conference on Acoustics Speech, and Signal Processing}{Proc. ICASSP}

\DefineJournalException{Proceedings of the IEEE International Conference on Acoustics Speech and Signal Processing}{Proc. ICASSP}

\DefineJournalException{Proceedings of the 2015 IEEE International Conference on Acoustics, Speech and Signal Processing}{Proc. ICASSP}

\DefineJournalException{Proceedings of the 2020 IEEE International Conference on Acoustics, Speech and Signal Processing}{Proc. ICASSP}
\DefineJournalException{Proceedings of the 2022 IEEE International Conference on Acoustics, Speech and Signal Processing}{Proc. ICASSP}

\DefineJournalException{Proceedings of the 2023 IEEE International Conference on Acoustics, Speech and Signal Processing}{Proc. ICASSP}

\DefineJournalException{Proceedings of the 19th International Society for Music Information Retrieval Conference}{Proc. ISMIR}

\DefineJournalException{Proceedings of the 21st International Society for Music Information Retrieval Conference}{Proc. ISMIR}

\DefineJournalException{Proceedings of the IEEE Workshop on Applications of Signal Processing to Audio and Acoustics}{Proc. WASPAA}

\DefineJournalException{Proceedings of the 2011 IEEE Workshop on Applications of Signal Processing to Audio and Acoustics}{Proc. WASPAA}

\DefineJournalException{Proceedings of the 2019 IEEE Workshop on Applications of Signal Processing to Audio and Acoustics}{Proc. WASPAA}

\DefineJournalException{Proceedings of the 30th European Signal Processing Conference}{Proc. EUSIPCO}

\DefineJournalException{Proceedings of the Annual Conference of the International Speech Communication Association}{Proc. Interspeech}

\DefineJournalException{Proceedings of the IEEE/CVF Conference on Computer Vision and Pattern Recognition}{Proc. CVPR}

\DefineJournalException{Proceedings of the 2007 International Conference on Independent Component Analysis and Signal Separation}{Proc. ICA}

\DefineJournalException{Proceedings of the 2018 International Conference on Latent Variable Analysis and Signal Separation}{Proc. LVA/ICA}

\DefineJournalException{Proceedings of the 33rd Conference on Neural Information Processing Systems}{Proc. NIPS}

\DefineJournalPartialAbbreviation{Neural}{Neur}

\DefineJournalException{Proceedings of the 34th International Conference on Machine Learning}{Proc. ICML}

\DefineJournalException{Proceedings of the 26th ACM International Conference on Multimedia}{Proc. ACM MM}

\DefineJournalException{The Journal of the Acoustical Society of America}{J. Acoust. Soc. Am.}

\newcommand{\sgn}{\operatorname{sgn}}
\renewcommand{\b}{\bfseries}
\def\rrparen{\mspace{1mu}\rrparenthesis}

\begin{document}
\receiveddate{5 September 2023}
\reviseddate{30 November 2023}
\accepteddate{30 November 2023}
\publisheddate{}
\currentdate{30 November 2023}
\doiinfo{}

\markboth{A Generalized Bandsplit Neural Network for Cinematic Audio Source Separation}{Watcharasupat {et al.}}

\title{A Generalized Bandsplit Neural Network for Cinematic Audio Source~Separation}

\author{%
    Karn N. Watcharasupat\authorrefmark{1*,2}, 
        Graduate~Student~Member,~IEEE,\authornote{Part of this work was done while K. N. Watcharasupat was supported by the AAUW International Fellowship\\from the American Association of University Women (AAUW) and the IEEE Signal Processing Society Scholarship Program.}\\ 
    Chih-Wei Wu\authorrefmark{1},
        Member,~IEEE, 
    Yiwei Ding\authorrefmark{2},
    Iroro Orife\authorrefmark{1},
        Member,~IEEE,\\
    Aaron J. Hipple\authorrefmark{1},
        Member,~IEEE,
    Phillip A. Williams\authorrefmark{1},
    Scott Kramer\authorrefmark{1},\\ 
    Alexander Lerch\authorrefmark{2}, Senior~Member,~IEEE, and
    William Wolcott\authorrefmark{1}
}
\affil{Netflix Inc., Los Gatos, CA, USA (* indicates work done during an internship)}
\affil{Music Informatics Group, Georgia Institute of Technology, Atlanta, GA, USA}
\corresp{Corresponding author: Karn N. Watcharasupat (email: kwatcharasupat@gatech.edu).}
% \authornote{\affilmark{*}Work done during an internship with Netflix, Inc.}

\begin{abstract}
Cinematic audio source separation is a relatively new subtask of audio source separation, with the aim of extracting the dialogue, music, and effects stems from their mixture. In this work, we developed a model generalizing the Bandsplit RNN for any complete or overcomplete partitions of the frequency axis. Psychoacoustically motivated frequency scales were used to inform the band definitions which are now defined with redundancy for more reliable feature extraction. A loss function motivated by the signal-to-noise ratio and the sparsity-promoting property of the 1-norm was proposed. We additionally exploit the information-sharing property of a common-encoder setup to reduce computational complexity during both training and inference, improve separation performance for hard-to-generalize classes of sounds, and allow flexibility during inference time with detachable decoders. Our best model sets the state of the art on the Divide and Remaster dataset with performance above the ideal ratio mask for the dialogue stem.
\end{abstract}


\maketitle

\sloppy

\section{INTRODUCTION}

Audio source separation refers to the task of separating an audio mixture into one or more of its constituent components. More formally, consider a set of source signals $\mathfrak{U}=\{\mathbf{u}_i \colon \mathbf{u}_i[n] \in \mathbb{R}^{D_i},\ n \in \llbracket0, M_i \rrparen \}$, where $i$ is the source index, $D_i$ is the number of channels in the $i$th source, $n$ is the sample index, $M_i$ is the number of samples in the $i$th source, and $\llbracket a, b \rrparen = \mathbb{Z} \cap [a, b)$. 
Not all of $\mathfrak{U}$ may be necessarily `desired'. 
The desired subset $\mathfrak{T} \subseteq \mathfrak{U}$ is often referred to as the set of `target' sources or stems, while the undesired subset $\mathfrak{N} = \mathfrak{U}\backslash\mathfrak{T}$ is often referred to as the set of `noise' sources. An input signal to a source separation (SS) system can usually be modeled as a mixing process 
\begin{equation}
    \mathbf{x} = \textstyle\sum_i \mathcal{T}_i(\mathbf{u}_i) \in \mathbb{R}^{C\times N}, \label{eq:ss}
\end{equation}
where $C$ is the number of channels in the mixture, $N$ is the number of samples in the mixture, and $\mathcal{T}_i\colon\mathbb{R}^{D_i \times M_i}\mapsto\mathbb{R}^{C\times N}$ is an audio signal transformation on the $i$th source. Some common operations represented by $\mathcal{T}_i$ are the identity transformation, which produces an instantaneous mixture often seen in synthetic data; a convolution, which produces a convolutive mixture often used to model a linear time-invariant (LTI) process; and a nonlinear transformation, often seen in music mixing process. The goal of an SS system is then to recover one, some, all, or composites of the elements of $\mathfrak{T}$, up to some allowable deformation~\cite{Vincent2006PerformanceSeparation, LeRoux2019SDRDone}. Note, however, that \eqref{eq:ss} does not take into account global nonlinear operations such as dynamic compression.

Composite targets are also often encountered in tasks such as music (e.g. the `accompaniment' stem) or cinematic SS (e.g. the `effects' stem), where the true number of component stems a composite target may contain can be fairly large. For simplicity concerning composite targets and multichannel sources, we will denote $\mathfrak{S}=\{ \mathbf{s}_i \colon {\mathbf{s}_i = \sum_j \mathcal{T}_j(\mathbf{u}_j)},\ {\mathbf{u}_j\in\mathfrak{T}},\ {\mathbf{s}_i[n] \in \mathbb{R}^{C}},\ {n \in \llbracket0, N \rrparen} \}$ as the set of `computational targets' of the algorithms. `Targets' in this manuscript will refer to $\mathfrak{S}$, as opposed to $\mathfrak{T}$.

Cinematic audio source separation (CASS) is a relatively new subtask of audio SS, 
most commonly concerned with extracting the dialogue, music, and effects stems from their mixture. 
Research traction in this new subtask can be credited to Petermann et al.~\cite{Petermann2023TheSoundtracks, Petermann2023TacklingSoundtracks} and the Cinematic Sound Demixing track of the Sound Demixing Challenge~\cite{Uhlich2023TheTrack}, introduced in 2023. 
While the setup of the task can be easily generalized from standard SS setups, the nature of cinematic audio poses a unique problem not commonly seen in speech or music SS. 
Specifically, CASS is closely related to universal audio SS, in which nearly the entire ontological categories of audio (speech, music, sound of things, and environmental sounds) must be all retrieved with equal or similar importance. Moreover, the ``music'' and ``effects'' stems can be very non-homogeneous. Music can consist of sound made by a very wide variety of acoustic, electronic, and synthetic musical instruments. More challengingly, the effects stem consists of anything that is \textit{not} speech or music, but also sometimes consists of sounds made by musical instruments in a non-musical context.

In this work, we adapted the Bandsplit RNN (BSRNN)~\cite{Luo2023MusicRNN} from the music SS task to the CASS task. 
In particular, we generalized the BSRNN architecture to potentially overlapping band definitions, introduced a loss function based on a combination of the 1-norm and the SNR loss, and modified the BSRNN from a set of single-stem models to a common-encoder system that can support any number of decoders. 
We further provide empirical results to demonstrate that the common-encoder setup provides superior results for hard-to-learn stems and allows generalization to previously untrained targets without the need for retraining the entire model. 
To the best of our knowledge, our proposed method\footnote{Replication code is available at \href{https://github.com/karnwatcharasupat/bandit}{github.com/karnwatcharasupat/bandit}.} is currently the state of the art on the Divide and Remaster (DnR) dataset~\cite{Petermann2023TheSoundtracks}.

\section{RELATED WORK}

Most early audio SS research was originally focused on a mixture of speech signals, particularly due to the reliance on statistical signal processing and latent variable models~\cite{Hyvarinen2000IndependentApplications}, which do not work well with more complex audio signals such as music or environmental sounds. Specifically, most early systems~\cite{Vincent2007FirstResults, Vincent2012TheChallenges, Stoter2018TheCampaign} assume an LTI mixing process, allowing for retrieval of target stems by means of filtering~\cite{Ozerov2010MultichannelSeparation}, matrix (pseudo-)inversion for (over)determined systems $C\ge D_i$~\cite{Ono2011StableTechnique}, or other similarity-based methods for underdetermined systems~\cite{Nguyen2020DirectionalMatrices}. These methods, however, often require fairly strong assumptions on the source signals such as statistical independence, stationarity, and/or sparsity. 

As computational hardware became more powerful, more computationally complex methods also became viable. This allowed for the relaxation of many statistical requirements placed on the signals in pursuit of more data-driven methods and the possibility of performing SS on nonlinear mixtures of highly correlated stems. Time-frequency (TF) masking, in particular, became the dominant method of source extraction in deep SS~\cite{Hershey2016DeepSeparation}. While this has led to major improvements in extracted audio quality, it came at the sacrifice of the interpretability once enjoyed in latent variable models. 

Denote $\mathbf{X} \in \mathbb{C}^{C \times F \times T}$ as the STFT of $\mathbf{x}$, where $F$ is the number of non-redundant frequency bins and $T$ is the number of time frames. Similarly, denote $\mathbf{S}_i$ as the STFT of the $i$th target source. Most masking SS systems use some form of $\hat{\mathbf{S}}_i = \mathbf{X} \circ \mathbf{M}$,
where $\hat{\mathbf{S}}_i $ is the estimate of $\mathbf{S}_i$, $\circ$ is elementwise multiplication with broadcasting, and $\mathbf{M}$ is the TF mask. Depending on the method, $\mathbf{M}$ may be binary, real-valued, or complex-valued, and has the same TF shape as $\mathbf{X}$, but may or may not be predicted separately for each channel. Although some works have generalized the masking operation to include additive components \cite{Sharma2023EgocentricSuppression} or more complex operations \cite{Watcharasupat2022End-to-EndSuppression}, direct masking still remains the most common method of source extraction, particularly due to its direct connection with time-variant convolution in the time domain. 
Many deep architectures have been proposed to predict the TF masks: Open-Unmix~\cite{Stoter2019Open-UnmixSeparation} used bidirectional LSTM  (BiLSTM) to obtain a magnitude mask; SepFormer~\cite{Subakan2021AttentionSeparation} applied a transformer to predict masks for speech separation, improving the performance while allowing parallel computing; (Conv-)TasNet~\cite{Luo2018TasNet:Separation, Luo2019Conv-TasNet:Separation} used masks on real-valued basis projections to allow real-time separation.

Despite the popularity of mask-based methods, several works have explored mask-free architectures. Wave-U-Net~\cite{Stoller2018Wave-U-Net:Separation} applies the U-Net structure to directly modify the mixture waveform. Built on Wave-U-Net, Demucs~\cite{Defossez2019MusicDomain} incorporates a BiLSTM at the bottleneck. Hybrid Demucs~\cite{Defossez2021HybridSeparation} extends the idea of combining time and frequency domains by applying two separated U-Nets for each domain with a shared bottleneck BiLSTM for cross-domain information fusion. Hybrid Transformer Demucs~\cite{Rouard2023HybridSeparation} further improves the performance by replacing the BiLSTM bottleneck with a transformer bottleneck. KUIELab-MDX-Net~\cite{Kim2021KUIELab-MDX-Net:Demixing} combines Demucs with a frequency-domain, U-Net-based architecture and uses a weighted average as the final output.

Under the definition in \eqref{eq:ss}, a number of non-generative audio enhancement tasks can also be considered special cases of audio SS, despite often not being actively thought of as one. Most non-generative implementations of noise suppression~\cite{Li2019MultichannelMemory, Li2020LearningMusic}, audio restoration~\cite{Deng2020ExploitingRestoration}, and dereverberation~\cite{Li2021ADecoupling, Fu2021DESNet:Separation} can be considered as an SS task with a noisy (and/or wet) mixture as input, and clean (and/or dry) target source as output. Dialogue enhancement often requires SS to extract the constituent stems before loudness adjustment is applied~\cite{Paulus2022SamplingSeparation}. Extraction of the dialogue stem in CASS, in particular, can be seen as closely related to the task of speech enhancement, while that of the music-and-effects (M\&E) stem can be seen as a speech suppression task.

Among deep learning-based SS models, several common meta-architectures exist.
Models such as Open-Unmix~\cite{Stoter2019Open-UnmixSeparation} and BSRNN~\cite{Luo2023MusicRNN} have one fully independent model for each stem, with no shared learnable layer. While this is very simple to train, fine-tune, and inference, the model suffers from the lack of information sharing between each stem-specific model. Adding additional stems to this system involves creating a completely separate network.

Some systems, such as Demucs~\cite{Defossez2021HybridSeparation, Rouard2023HybridSeparation} and Conv-TasNet~\cite{Luo2019Conv-TasNet:Separation}, use one shared model for all stems. This means that training and inference must happen for all stems at the same time. This setup is perhaps the most beneficial in terms of information sharing, but it is also difficult to understand the flow of information within the system, as all intermediate representations are entangled up until the last layer. It can also be very difficult to add an additional stem to the model, as it is not trivial to decide which part of the model parameters may be safe to freeze or unfreeze.
\section{PROPOSED METHOD} \label{sec:pm}

\begin{figure}
    \centering
    \includegraphics[width=0.65\columnwidth]{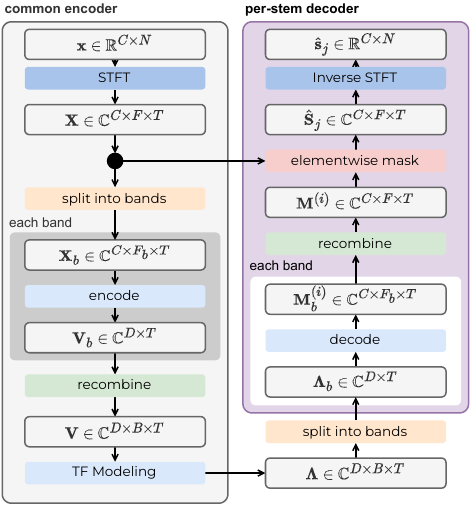}
    % \vspace{-5pt}
    \caption{Overview of the proposed model architecture, Bandit.}
    \label{fig:model}
\end{figure}

Our proposed method builds upon the BSRNN model proposed in \cite{Luo2023MusicRNN}. BSRNN itself is related to works that split the frequency bands into several different groups~\cite{Takahashi2021DenselyTasks, Wang2023TF-GridNet:Separation}, and those that apply multi-path recurrent networks to deal with long sequences~\cite{Luo2020Dual-PathSeparation, Kinoshita2020Multi-pathSeparation}. The original BSRNN is very similar in structure to our proposed model in \Cref{fig:model}, but with a separate model per stem. Each BSRNN model consists of a bandsplitting module, a TF modeling module, and a mask estimator. The bandsplitting module in~\cite{Luo2023MusicRNN} partitions an input spectrogram along its frequency axis into $B$ disjoint ``bands'', then, in parallel, performs a normalization and an affine transformation for each band. Each affine transformation contains the same number of $D$ output neurons. The TF module consists of a stack of bidirectional RNNs operating alternatingly along the time and band axes of the feature map. In~\cite{Luo2023MusicRNN}, this consists of a stack of 12 pairs of residually-connected BiLSTMs. Finally, the mask estimation module consists of $B$ parallel feedforward modules which produce $B$ bandwise complex-valued masks.

% Our contributions lie in the investigation of the common-encoder architecture, band definitions, loss functions, and generalizability. 
The overview of the proposed model is shown in \Cref{fig:model}. For clarity, BSRNN will only refer to the original model in~\cite{Luo2023MusicRNN}. Our proposed model will be referred to as ``BandIt''\footnote{From \textbf{band}spl\textbf{it}, and a reference to the multi-armed bandit problem.}.

\subsection{Common Encoder} \label{ssec:ce}

In this work, we propose to use a common-encoder multiple-decoder system. By treating multi-stem SS as a multi-task problem, this is akin to hard parameter sharing.
% Note that we use the term encoder and decoder very loosely here since our system does not have a true ``bottleneck''. In these systems, we have a single shared encoder (i.e. feature extractor) to produce an intermediate feature. 
This system allows information sharing to occur freely in the encoder section, but not in the decoder. It is likely that this can improve the information efficiency, and generalizability of the model~\cite{Ravanbakhsh2017EquivarianceParameter-Sharing, Cheng2018LearningSearch}. A downside of this system is that adding a new decoder may or may not require the encoder to be retrained, depending on the generalizability of the feature maps after the initial training with the original set of stems. 

In addition to the potential information theoretic benefits, the common-encoder structure offers a more practical benefit in terms of the computational requirements. Training using the common encoder system can reduce the amount of parameters needed considerably, and thus reduce memory and hardware requirements. 
Additionally, in the case where not all decoders can be trained concurrently, simultaneous training can still be approximated by only attaching a subset of the decoders at each optimization step and alternating over them. Finally, this allows an arbitrary number of decoders to be attached and detached as needed during inference. 

As seen in \Cref{fig:model}, BSRNN can be modified into a common-encoder BandIt by sharing the all modules up to the TF modeling module and only splitting into stem-specific modules at the mask estimator section. Of course, many other possible points of splitting exist; we chose to split only after the TF modeling module in order to force it to learn a common representation that will work for all three stems.

\begin{figure*}[tb]
    \centering
    \includegraphics[width=0.84\textwidth]{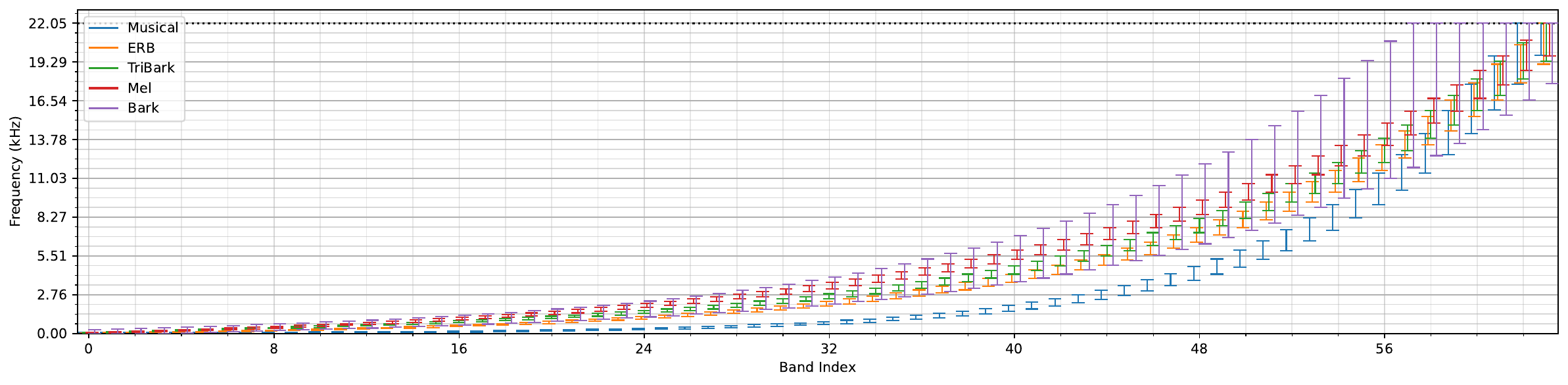}
    \vspace{-5pt}
    \caption{Frequency ranges of each band, by band type, for a 64-band setup with a sampling rate of \SI{44.1}{\kilo\hertz} and an FFT size of \SI{2048}{samples}.}
    \label{fig:bands}
\end{figure*}

\subsection{Bandsplit Module} \label{ss:band}

The original definition of the bands in BSRNN has two clear attributes: (A1) the bandwidth in \si{\hertz} generally increases with its constituent frequencies, and (A2) the number of bands is high in regions where the sources of a stem typically are most active in. From a data compression perspective, this translates to the assumption that (B1) information content per \si{\hertz} decreases with increasing frequency, and (B2) information content is positively correlated to source activity. Both ``priors'' may seem trivial. However, the implementation can be tricky as we will discuss below.

In~\cite{Luo2023MusicRNN}, band definitions were mostly handcrafted for each stem. This potentially limits the generalizability of the model and makes architecture design difficult when dealing with stems with unpredictable, non-homogeneous content such as the ``other'' stem in MUSDB18~\cite{musdb18} and the effects stem in cinematic audio. In other words, the model is prone to prior mismatch when dealing with very diverse content. Moreover, the band definitions in~\cite{Luo2023MusicRNN} are all disjoint, i.e., each frequency bin is allocated to only one band. From a system reliability perspective, this means that the very first layer of BSRNN already has no redundancy provisioned; any loss of information occurring during the first affine transformation cannot be recovered by other parallel affine modules. This also disproportionally affects semantic structures (i.e. the ``blobs'' in spectrogram) that are located around the band edges, since they will be broken up into two disjoint bands, resulting in neither of which being able to encode their information well.

To deal with these issues, we limit the prior assumption to only (B1), turning to psychoacoustically motivated band definitions in lieu of handcrafting. Additionally, we propose to add redundancy to the bandsplitting process in an attempt to reduce the amount of early information loss. Specifically, we will investigate five different band definitions based on four frequency scales with psychoacoustic motivations, namely, the mel scale, the equivalent rectangular band (ERB) scale, the Bark scale, and the 12-tone equal temperament (12-TET) Western musical scale. Note that we do not directly use the bandwidths associated with the ERB and the Bark scale, but rather take the scale value as a rough approximation of the number of critical bands below it.

For all scale-filterbank combinations, the proposed splitting process is as follows. The minimum scale value $z^{\text{min}}$ and the maximum $z^{\text{max}}$ were computed. For all scales, $z^{\text{max}}$ is given by $z(0.5 f_\text{s})$, where $z\colon \mathbb{R}_0^+ \mapsto \mathbb{R}$ is the mapping function from Hz to the scale's unit,  and $f_\text{s}$ is the sampling rate in Hz. For the mel, ERB, and Bark scales, $z^{\text{min}} = 0$. The $z^{\text{min}}$ musical scale will be detailed later. For $B$ bands, the center frequencies, in each respective scale, are given by 
\begin{equation}
    \zeta_n = {z(0.5f_\text{s})}\cdot \left(n+1\right) / ({B + 2}).
\end{equation}
The frequency weights $\mathbf{W} \in [0, 1]^{B \times F}$ are then computed using a filterbank of choice, 
% with the $b$th filter having band edges $p_{n-1}$ and $p_{n+1}$, 
and its weights normalized so that $\sum_{b} \mathbf{W}[b,f] = 1, \forall f \in \llbracket 0, F \rrparen$. 
Using the filterbank values, the band definitions are then created using a simple binarization criterion
\begin{align}
    \mathfrak{F}_b = \{f \in \llbracket 0, F  \rrparen \colon \mathbf{W}[b, f] > 0\},\ \forall b \in  \llbracket 0, B \rrparen.
\end{align}
We then define a subband $\mathbf{X}_b \in \mathbb{C}^{C \times F_b \times T}$ of $\mathbf{X}$ such that
\begin{equation}
    \mathbf{X}_b = \mathbf{X}[:_c, \mathfrak{F}_b, :_t],\ \forall b \in  \llbracket 0, B \rrparen.
\end{equation}
The scales and the filterbanks used are detailed as follows, and visualized in \Cref{fig:bands}.

\subsubsection{Mel Scale}

The mel scale is one of the most used scales for the calculation of input features, such as the (log-)mel spectrogram and the mel-frequency cepstrum coefficients, for many audio tasks in machine learning and information retrieval. It is a measure of \textit{tone height} \cite{Lerch2023TonalAnalysis}.
In this work, we use the mel scale given in~\cite[p.128]{OShaughnessy2000Hearing}, where
\begin{equation}
    z_{\text{mel}}(f)
    = 2595 \log_{10}\left(1 + {f}/{700}\right).
\end{equation}
The filterbank used is comprised of triangular-shaped filters with the $b$th filter having band edges $\zeta_{b-1}$ and $\zeta_{b+1}$, similar to the implementations in librosa \cite{McFee2015Librosa:Python} and PyTorch \cite{Paszke2019PyTorch:Library}.

\subsubsection{Bark scale}

The Bark scale~\cite{Zwicker1961SubdivisionFrequenzgruppen} ``relates acoustical frequency to perceptual frequency resolution, in
which one Bark covers one critical bandwidth \cite[p.128]{OShaughnessy2000Hearing}''. Also known as the \textit{critical band rate}, the Bark scale is constructed from the bandwidth of measured frequency groups \cite{Lerch2023TonalAnalysis}. 
% It is most commonly used in relation to the concept of critical bands of human hearing. 
Unlike the mel scale, the Bark scale is more concerned with the widths of the critical bands than the center frequencies themselves.  
In this work, we use the approximation~\cite{Wang1992AnCoders} given by
\begin{equation}
    z_{\text{bark}}(f)
    = 6 \sinh^{-1}\left({f}/{600}\right).
\end{equation}
For the Bark scale, we experimented with two filterbanks. One is a Bark filterbank implementation provided by Spafe~\cite{Malek2023Spafe:Extraction}, and another is a simple triangular filterbank similar to the mel and ERB scales. The former will be referred to as the ``Bark'' bands, and the latter as ``TriBark''.

\subsubsection{Equivalent Rectangular Bandwidth Scale}

The equivalent rectangular bandwidth (ERB) was designed with a similar motivation to the Bark scale. The ERB is an approximation of the bandwidth of the human auditory filter at a given frequency. The ERB scale is a related scale that computes the number of ERBs below a certain frequency. The ERB scale can be modeled as~\cite{Glasberg1990DerivationData}
\begin{equation}
    z_{\text{erb}}(f)
    = \ln\left(1 + \num{4.37e-3} f\right)/(24.7 \cdot \num{4.37e-3}). 
\end{equation}
The filterbank is computed similarly to that of the mel scale.

\subsubsection{12-TET Western Musical Scale}

The 12-TET scale is the most common form of Western musical scale used today. 
Using a reference frequency of $f_\text{ref} =\SI{440}{\hertz}$, the unrounded MIDI note number of a particular pitch can be represented by
\begin{equation}
    \tilde{z}_{\text{mus}}(f) = 69 + 12\log_{2}\left({f}/{f_\text{ref}}\right).
\end{equation}
Crucially, scaling a frequency by a factor of $k$, always lead to a constant change in this scale by $12\log_2 k$, i.e.,
\begin{equation}
\tilde{z}_{\text{mus}}(kf) = \tilde{z}_{\text{mus}}(f) + 12\log_2 k.
\end{equation}
This ensures that the $k$th harmonic of a sound is always $12\log_2 k$ note numbers away from its fundamental, regardless of the fundamental pitch --- a property that mel, ERB, and Bark scales do not enjoy. 
% This has a property of converting the equal frequency ratio to a constant shift representation, which is in line with both the physical acoustics of many natural occurring sounds, and the musical properties arising from these frequency ratio. 
% The musical scale is closely related to the constant-Q transfrom (CQT) and the fractional octave filters. 
In practice, since $\tilde{z}_{\text{mus}}(f\to 0^+) \to -\infty^{+}$, we instead set scale value as
\begin{equation}
    z_{\text{mus}}(f) = \max\left[z^{\text{min}}_{\text{mus}}, \tilde{z}_{\text{mus}}\left(f\right)\right],
\end{equation}
where $z^{\text{min}}_{\text{mus}} = \tilde{z}_{\text{mus}}\left({f_\text{s}}/{N_\text{FFT}}\right)$, and $N_\text{FFT}$ is the FFT size. 

In this work, the filterbank for the musical scale is implemented using rectangular filters with the $b$th filter having band edges $\zeta_{b-1}$ and $\zeta_{b+1}$. All filters, except for the lowest and highest bands, have the same bandwidth in cents, before being discretized to match FFT bins. For brevity, we will refer to this band type simply as ``musical''. 
A comparison of the five proposed band definitions is shown in \Cref{fig:bands}.

\begin{figure}[t]
    \centering
    \includegraphics[width=0.75\columnwidth]{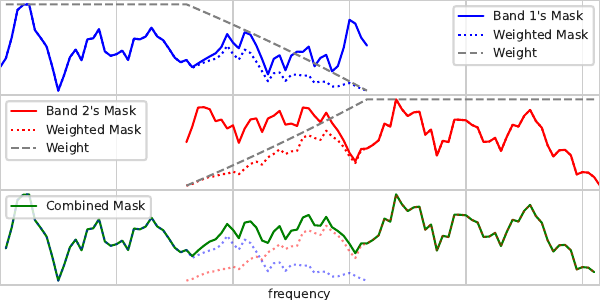}
    \vspace{-5pt}
    \caption{A simplified illustration of overlapping mask recombination.}
    \label{fig:mask}
\end{figure}

\subsection{Bandwise Feature Embedding}
After splitting, each of the subbands is viewed as a real-valued tensor in $\mathbb{R}^{2CF \times T}$ by collapsing the channel and frequency axes and then concatenating its real and imaginary parts. As with BSRNN \cite[Fig. 1b]{Luo2023MusicRNN}, each band is passed through a layer normalization and an affine transformation with $D=\num{128}$ output units along the pseudo-frequency axis. The feature embedding process is denoted by $\mathcal{P}_b \colon \mathbb{C}^{C\times F_b \times T} \mapsto \mathbb{R}^{D \times T}$. The bandwise feature tensors are then stacked to obtain the full-band feature tensor $\mathbf{V} \in \mathbb{R}^{D\times B \times T}$ such that ${\mathbf{V}[:, b, :]} = \mathcal{P}_b(\mathbf{X}_b)$, $\forall b \in  \llbracket 0, B \rrparen$.
Except for the Bark model, the feature embedding module accounts for approximately \SI{600}{\kilo{}} parameters in a 64-band setup.

\subsection{Time Frequency Modeling}

As with BSRNN \cite[Fig. 1c]{Luo2023MusicRNN}, the feature tensor $\mathbf{V}$ is passed through a series of residual recurrent neural networks (RNNs) with affine projection, alternating its operation between the time and frequency axes. In this work, we reduced the number of residual RNN pairs from 12 to 8 and also opted to use Gated Recurrent Units (GRUs) instead of Long-Short Term Memory (LSTM) units as the RNN backbone. As with \cite{Luo2023MusicRNN}, each RNN has $2D$ hidden units.
% Our preliminary experiments have found that we can reduce the total number of units and get better parameter efficiency with GRU instead of LSTM while maintaining similar separation performance.
The overall operation of this module is represented by the transformation $\mathcal{R} \colon \mathbb{R}^{D \times B \times T} \mapsto \mathbb{R}^{D \times B \times T}$ to obtain the output $\mathbf{\Lambda} = \mathcal{R} \left(\mathbf{V}\right) \in \mathbb{R}^{D \times B \times T}$.
TF modeling with 8 residual GRU pairs accounts for \SI{10.5}{\mega{}} trainable parameters\footnote{Due to the computational complexity of backpropagation through time with long sequences, we experimented with replacing the RNNs with transformer encoders or convolutional layers. With similar numbers of parameters and all else being equal, these were not able to match the performance of an RNN-based module.}.

\subsection{Overlapping Mask Estimation and Recombination}

At this stage, the shared feature $\mathbf{\Lambda}$ is passed to a separate mask estimator for each stem. The internal implementation of the mask estimation module is identical to that of the original BSRNN. The overall operation of this module is represented by $\mathcal{Q}^{(i)}_{b, \text{re}}, \mathcal{Q}^{(i)}_{b, \text{im}} \colon \mathbb{R}^{D \times B \times T} \mapsto \mathbb{R}^{C \times F_b \times T}$ to obtain the bandwise mask
\begin{equation}
    \mathbf{M}^{(i)}_b = \mathcal{Q}^{(i)}_{b, \text{re}}(\mathbf{\Lambda}_b) + \jmath \mathcal{Q}^{(i)}_{b, \text{im}}(\mathbf{\Lambda}_b) \in  \mathbb{C}^{C \times F_b \times T}.
\end{equation}
With overlapping bands, however, the full-band mask can no longer be trivially obtained using stacking. We used weighted recombination to obtain $\mathbf{M}^{(i)}\in\mathbb{C}^{C \times F \times T}$, such that
\begin{equation}
    \mathbf{M}^{(i)}[c, f, t] = \textstyle\sum_b \mathbf{W}_b[f] \cdot \mathbf{M}^{(i)}_b[c, f - \min \mathfrak{F}_b , t] 
\end{equation}
A simplified illustration with two bands is shown in \Cref{fig:mask}. Note that while $\mathbf{W}_b$ is used as the recombination weight, it is possible to not use any weight as $\mathbf{W}_b$ or more appropriate weights can be learned by the model and be absorbed into $\mathbf{M}^{(i)}_b$. In other words, the role of $\mathbf{W}_b$ in the mask estimation module is more of an initialization than a fixed parameter. Except for the Bark model with very wide bandwidths thus a higher number of parameters, the mask estimation module accounts for roughly \SI{25}{\mega{}} parameters in a 64-band setup.\footnote{We have also attempted a combination of multiplicative and additive masks in this work. However, we found that the inclusion of the additive mask did not lead to any appreciable improvement. We hypothesize that the channel capacity of the model is simply insufficient to reconstruct a sufficiently good full-resolution additive spectrogram, as a non-zero additive will only lead to more artifacts.}

\subsection{Loss function}\label{ssec:loss}

We initially experimented with the loss function originally used in~\cite{Luo2023MusicRNN, Wang2021OnSeparation}, whose stem-wise contribution is given by
\begin{align}
    \mathcal{L}^{(i)}_p     = 
    \|\hat{\mathbf{s}}_i - \mathbf{s}_i\|_p + 
    \|\Re [\hat{\mathbf{S}}_i - \mathbf{S}_i]\|_p + \|\Im[\hat{\mathbf{S}}_i - \mathbf{S}_i]\|_p,
\end{align}
and $p = 1$. While calculating the loss for the real and imaginary parts separately may seem like a somewhat inelegant approximation, there is a desirable gradient behavior that justifies doing so over calculating a norm of complex differences. 
Consider $\mathbf{y} = \mathbf{u}+\jmath\mathbf{v}$ and $\hat{\mathbf{y}} =\hat{\mathbf{u}}+\jmath\hat{\mathbf{v}}$.
The gradient of the 1-norm of a complex difference vector gives
\begin{align}
    \partial\|\hat{\mathbf{y}}-\mathbf{y}\|_1
    &= \sum\mathop{}_i \dfrac{(\hat{u}_i - u_i)\partial \hat{u}_i + (\hat{v}_i - v_i)\partial \hat{v}_i}{\sqrt{(\hat{u}_i - u_i)^2 + (\hat{v}_i - v_i)^2}}.
\end{align}
This indicates that the gradient $\partial \hat{u}_i$ will be scaled down if the error on $\hat{v}_i$ is high and vice versa, diluting the sparseness-encouraging property of a $1$-norm. On the other hand, treating the real and imaginary parts separately yields
\begin{align}
    &\partial\left(\|\hat{\mathbf{u}}-\mathbf{u}\|_1+\|\hat{\mathbf{v}}-\mathbf{v}\|_1\right)\nonumber
    \\&\qquad
    = \sum\mathop{}_i \sgn(\hat{u}_i - u_i)\partial \hat{u}_i + \sgn(\hat{v}_i - v_i)\partial \hat{v}_i,
\end{align}
which enjoys the same sparsity benefit of a $1$-norm for real-valued differences.

Both acoustically and perceptually, however, the magnitudes of both the time-domain signal and the STFT follow a logarithmic scale. Each of the stems can also have very different energies due to foreground (e.g., dialogue) sources conventionally being mixed louder than background (e.g., music and effects) sources. Inspired by the success of negative signal-to-noise ratio (SNR) as a loss function, we experimented with a generalization to a $p$-norm that tackles both of these issues, i.e.,
\begin{equation}
    \mathcal{D}_p(\hat{\mathbf{y}}; \mathbf{y})
    = 10\log_{10} \left[(\|\hat{\mathbf{y}} - \mathbf{y}\|_p^p  + \epsilon)/(\|\mathbf{y}\|_p^p + \epsilon)\right],
\end{equation}
where $\epsilon$ is a stabilizing constant, setting the minimum of the distance to $-10\log_{10}(\epsilon^{-1}\|\mathbf{y}\|_p^p + 1)$, which is numerically stable for $\epsilon \not\ll \|\mathbf{y}\|_p^p$. In this work, we set $\epsilon = \num{e-3}$. Analyzing the differential of $\mathcal{D}_p$ gives
\begin{equation}
    \partial \mathcal{D}_p
    = \log_{10}(e^{10}) \cdot(\|\hat{\mathbf{y}}-\mathbf{y}\|_p^p+\epsilon)^{-1} \cdot \partial \|\hat{\mathbf{y}}-\mathbf{y}\|_p^p
\end{equation}
which allows the model to take smaller updates when it is less confident, and larger updates once it is more confident. Gradient explosion is prevented by $\epsilon$ since the magnitude of the gradients cannot rapidly increase once $\|\hat{\mathbf{y}}-\mathbf{y}\|_p \ll \epsilon$.  
Note also the importance of $p$ on the differential, since
\begin{align}
    \partial \mathcal{D}_1(\hat{\mathbf{y}}; \mathbf{y}) &= \dfrac{\log_{10}(e^{10})}{\|\hat{\mathbf{y}}-\mathbf{y}\|_1+\epsilon} \sum_{i} \sgn(\hat{y}_i - y_i) \cdot \partial \hat{y}_i, \\
    \partial \mathcal{D}_2(\hat{\mathbf{y}}; \mathbf{y}) &= \dfrac{2\log_{10}(e^{10})}{\|\hat{\mathbf{y}}-\mathbf{y}\|_2^2+\epsilon} \sum_{i} (\hat{y}_i - y_i) \cdot \partial \hat{y}_i .
\end{align}
While both differentials were globally modulated by the inverse norm of the error in both cases, $\partial \mathcal{D}_2$ is more prone to outliers in the early stage of training and to the vanishing gradient problem in the later stage due to the elementwise multiplier of $\partial\hat{y}_i$ being dependent on the elementwise error magnitude. On the other hand, the elementwise multiplier in $\partial \mathcal{D}_1$ only depends on the sign of the error and thus does not suffer from either problem.
Combining $\mathcal{D}_1$ with the original loss function gives
\begin{align}
    \mathcal{L}_\text{proposed}
    = \mathcal{D}_1(\hat{\mathbf{s}}; \mathbf{s})
    + \mathcal{D}_1(\Re\hat{\mathbf{S}}; \Re\mathbf{S})
    + \mathcal{D}_1(\Im\hat{\mathbf{S}}; \Im\mathbf{S}),\label{eq:loss}
\end{align}
which we will refer to as the proposed ``L1SNR'' loss. In practice, care must be taken to ensure that the DFT used in the STFT is normalized such that all loss terms are on a similar scale, or appropriate weightings should be used.

\section{EXPERIMENTAL SETUP}

\subsection{Dataset}

Most of the experiments in this work will focus on the Divide and Remaster (DnR) dataset~\cite{Petermann2023TheSoundtracks}. The DnR dataset is a three-stem dataset consisting of the dialogue, music, and effects stems. Each track is \SI{60}{\second} long, single-channel, and provided at two sample rates of \SI{16}{\kilo\hertz} and \SI{44.1}{\kilo\hertz}. In this work, we will only focus on the high-fidelity sample rate. 

The dialogue data were obtained from LibriVox, an English-only audiobook reading. Music data were taken from the Free Music Archive (FMA). Foreground and background effects data were taken from FSD50k. As mentioned in CDX~\cite{Uhlich2023TheTrack}, the dialogue data is not as diverse as real motion picture audio, due to the lack of emotional and linguistic diversity. Dialogue data diversity is particularly an issue when seeking high-fidelity speech sampled at \SI{44.1}{\kilo\hertz} and above; our own initial attempt to augment the DnR dataset with more languages and emotions required unexpectedly significant effort and was deferred to future work.

\subsection{Chunking}

Since each track of the DnR dataset is relatively long, the tracks were chunked during training and inference. During training, random \SI{6}{\second} chunks of the tracks are drawn on the fly. During validation, chunks were drawn exhaustively with a length of \SI{6}{\second} and a hop size of \SI{1}{\second}. During testing, we chunk the full signal into \SI{6}{\second} chunks with a hop size of \SI{0.5}{\second}. Inference is performed independently on each chunk before they are recombined with Hann-windowed overlap-add. The \SI{6}{\second} chunk size was originally chosen for compatibility with the original BSRNN implementation. It was also the largest chunk size we could fit into an NVIDIA A10G GPU with a per-GPU batch size of at least two, as a per-GPU batch size of one caused significant instability during backpropagation.

\subsection{Training}

Unless otherwise stated, all models were trained using an Adam optimizer for 100 epochs. The learning rate is initialized to \num{e-3} with a decay factor of \num{0.98} every two epochs. Norm-based gradient clipping was additionally enabled with a threshold of 5. Each training epoch is set to \SI{20}{\kilo{}} samples regardless of the dataset size.

As additional points of comparison, we trained our adaptation of the Hybrid Demucs~\cite{Defossez2021HybridSeparation} and Open-Unmix (\texttt{umxhq}-like)~\cite{Stoter2019Open-UnmixSeparation} for the 3-stem problem. The loss function for each model follows that of the respective original paper, while the data processing is identical to our proposed method. BandIt, BSRNN, and Demucs models were trained on a g5.48xlarge Amazon EC2 instance with 8 NVIDIA A10G GPUs (24 GB each). Training was done with PyTorch Lightning using a distributed data-parallel strategy with a batch size of 2 per GPU. Open-Unmix model was trained on a g4dn.4xlarge Amazon EC2 instance with a single NVIDIA T4 GPU (16 GB) with a batch size of 16. BandIt models each took roughly 1.5 days to complete 100 epochs of training. 

\subsection{Metrics}

In this work, we report the signal-to-noise ratio (SNR) and scale-invariant SNR (SI-SNR)~\cite{LeRoux2019SDRDone}. Note that the commonly reported signal-to-distortion ratio (SDR) and its scale-invariant counterpart (SI-SDR) are mathematically identical to SNR and SI-SNR, respectively, when the appropriate version of SDR is used~\cite{LeRoux2019SDRDone}. To avoid ambiguity, we will simply report the ``SNR'' and the ``SI-SNR''.

\begin{table*}[!t]
    \setlength{\tabcolsep}{3pt}
    \caption{Model performance on the DnR test set. Floating-point operation count is based on 6-second input at 44.1 kHz}
    \label{tab:results}
    \footnotesize
    \centering
    \begin{tabularx}{\linewidth}{
        lllX
        rS[table-format=5.1]
        S[table-format=-2.1]S[table-format=-2.1]
        S[table-format=-2.1]S[table-format=-2.1]
        S[table-format=-2.1]S[table-format=-2.1]
        S[table-format=-2.1]S[table-format=-2.1]
    }
    \toprule
    \multicolumn{4}{l}{\b Model} &&
    & \multicolumn{2}{r}{\b Dialogue} 
    & \multicolumn{2}{r}{\b Music} 
    & \multicolumn{2}{r}{\b Effects}  
    & \multicolumn{2}{r}{\b Averaged}  \\
    \cmidrule(r){1-4}
    \cmidrule(l){7-8} 
    \cmidrule(l){9-10} 
    \cmidrule(l){11-12} 
    \cmidrule(l){13-14} 
    \b Backbone & \b Encoder & \b Bands &  \b Loss & \b Params. &  \multicolumn{1}{r}{\b GFlops} &
        % DnR
        \multicolumn{1}{r}{\b SNR} & \multicolumn{1}{r}{\b SI-SNR} &
        \multicolumn{1}{r}{\b SNR} & \multicolumn{1}{r}{\b SI-SNR} & 
        \multicolumn{1}{r}{\b SNR} & \multicolumn{1}{r}{\b SI-SNR} & 
        \multicolumn{1}{r}{\b SNR} & \multicolumn{1}{r}{\b SI-SNR}  \\
    \midrule
    BSRNN-LSTM12 & Separate & Vocals V7 & T+RITF L1   & \SI{77.4}{\mega{}}  & 1386.5 &
        14.2 & 14.0 &
        6.3 & 5.2 & 
        7.0 & 5.9 &
        9.2 & 8.4\\
    BSRNN-GRU8 & Separate & Vocals V7 & T+RITF L1   & \SI{47.4}{\mega{}}  & 714.5 &
        14.0 & 13.9 & 
        6.4 & 5.2 & 
        7.2 & 6.2 & 
        9.2 & 8.4           \\
    \midrule
    BandIt & Shared & Vocals V7 
    & T+RITF L1 & \SI{25.7}{\mega{}}     & 243.2 &
        13.3 & 13.0 &
        6.4 & 5.3 & 
        7.8 & 6.9 & 
        9.2 & 8.4 \\
    && Vocals V7 & T+RITF MSE & \SI{25.7}{\mega{}}  & 243.2 &
        12.5 & 12.2 &
        5.5 & 4.1 & 
        7.0 & 6.0 &
        8.3 & 7.4 \\
    && Vocals V7 & T+RITF L1SNR & \SI{25.7}{\mega{}}  & 243.2 &
        14.2 & 14.0 &
        7.2 & 6.3 &
        8.5 & 7.8 &
        10.0 & 9.4\\
    && Vocals V7 & T+RITF L2SNR & \SI{25.7}{\mega{}}  & 243.2 &
        13.5 & 13.3 &
        6.5 & 5.4 &
        7.9 & 7.1 &
        9.3 & 8.6 \\
    \cmidrule{3-14}
    & &
     Bark 48 & T+RITF L1SNR    & \SI{64.5}{\mega{}} & 290.6 &
        14.1 & 14.0 & 
        7.3 & 6.3 & 
        8.6 & 7.8 &
        10.0 & 9.4 \\
    & &     Mel 48 & T+RITF L1SNR    & \SI{32.8}{\mega{}} & 274.3 &
        14.5 & 14.3 &
        7.5 & 6.6 &
        8.8 & 8.1 & 
        10.3 & 9.7 \\
    & & TriBark 48 & T+RITF L1SNR    & \SI{32.7}{\mega{}} & 274.2 &
        14.6 & 14.5 & 
        7.6 & 6.7 & 
        8.9 & 8.2 &
        10.4 & 9.8 \\
     & & ERB 48 & T+RITF L1SNR    & \SI{32.6}{\mega{}} & 274.2 &
        14.6 & 14.4 &
        7.7 & 6.8 &
        8.9 & 8.5 & 
        10.4 & 9.8 \\
     & & Music 48 & T+RITF L1SNR    & \SI{33.5}{\mega{}} & 274.7 & 
        14.8 & 14.6 &
        7.9 & 7.1 & 
        9.2 & 8.5 &
        10.6 & 10.1 \\
    \cmidrule{3-14}
    & & Mel 64 & T+RITF L1SNR    & \SI{36.1}{\mega{}} & 363.6 &
        14.8 & 14.7 &
        7.9 & 7.1 & 
        9.1 & 8.5 &
        10.6 & 10.1\\
     & & TriBark 64 & T+RITF L1SNR   & \SI{36.0}{\mega{}} & 363.5 &
        14.8 & 14.7 &
        7.9 & 7.1 & 
        9.1 & 8.4 &
        10.6 & 10.1 \\
     & & ERB 64 & T+RITF L1SNR    & \SI{36.0}{\mega{}} & 363.5 &
        15.0 & \b 14.9 &
        8.0 & 7.2 & 
        9.2 & 8.6 &
        10.8 & 10.2     \\
     & & Bark 64 & T+RITF L1SNR   & \SI{82.6}{\mega{}} & 387.6 &
        15.0 & \b 14.9 &
        8.1 & 7.3 & 
        \b 9.3 & 8.6 &
        10.6 & \b 10.3 \\
    & & Music 64 & T+RITF L1SNR    & \SI{37.0}{\mega{}} & 364.1 & 
        \b 15.1 & \b 14.9 &
        \b 8.2 & \b 7.4 &
        \b 9.3 & \b 8.7 &
        \b 10.9 & \b 10.3 \\
    \midrule
    BandIt+ & Shared & Music 64 & T+RITF L1SNR    & \SI{37.0}{\mega{}} & 364.1 & 
         15.7 &  15.6 &
         8.7 &  8.0 &
         9.8 &  8.2 &
         11.4 &  10.9 \\
    \midrule
    \multicolumn{3}{l}{Open-Unmix (\texttt{umxhq})} & TF Mag. MSE  & \SI{22.1}{\mega{}}& 5.7 &
        11.6 & 11.3 & 
        4.9 & 3.2 & 
        5.8 & 4.4 & 
        7.4 & 6.3  \\
    \multicolumn{3}{l}{MRX$^\triangle$} & Time SI-SDR & N/R & \multicolumn{1}{r}{N/R}&
        \multicolumn{1}{r}{---} & 12.3 &
        \multicolumn{1}{r}{---} & 4.2 &
        \multicolumn{1}{r}{---} & 5.7 &
        \multicolumn{1}{r}{---} & 7.4 \\
    \multicolumn{3}{l}{MRX-C$^\triangle$}& Time SI-SDR & N/R & \multicolumn{1}{r}{N/R} &
        \multicolumn{1}{r}{---}& 12.6 &
        \multicolumn{1}{r}{---} & 4.6 &
        \multicolumn{1}{r}{---} & 6.1 &
        \multicolumn{1}{r}{---} & 7.8 \\
    \multicolumn{3}{l}{Hybrid Demucs (v3)} & Time L1 & \SI{83.6}{\mega{}} & 85.0  & 
        13.6 & 13.4 &
        6.0 & 4.7 &
        7.2 & 6.1 &
        8.9 & 8.1 \\
    \midrule
    \multicolumn{3}{l}{\textit{Mixture}} & \multicolumn{1}{c}{---} & \multicolumn{1}{r}{---} & \multicolumn{1}{r}{---} &
        1.0 & 1.0 &
        -6.8 & -6.8 &
        -5.0 & -5.0 &
        -3.6 & -3.6 \\
    \multicolumn{3}{l}{\textit{Ideal Ratio Mask}} & \multicolumn{1}{c}{---} & \multicolumn{1}{r}{---}& \multicolumn{1}{r}{---} &
        14.4 & 14.6 & 
        9.0 & 8.4 & 
        11.0 & 10.7 & 
        11.5 & 11.2  \\
    \multicolumn{3}{l}{\textit{Phase Sensitive Filter}}& \multicolumn{1}{c}{---}&\multicolumn{1}{r}{---} & \multicolumn{1}{r}{---} &
        18.5 & 18.4 & 
        12.9 & 12.7 & 
        15.0 & 14.8 & 
        15.4 & 15.3  \\
    \bottomrule
    \end{tabularx} 
\end{table*}

\section{RESULTS AND DISCUSSION}

The main experimental results (\S V-A through \S V-D) are presented in \Cref{tab:results}. In addition to our proposed method, we trained and evaluated our own baselines with Open-Unmix~\cite{Stoter2019Open-UnmixSeparation} and Hybrid Demucs (a.k.a. Demucs v3)~\cite{Defossez2021HybridSeparation} on DnR. Results for the MRX and MRX-C models are reproduced as-is from~\cite{Petermann2023TacklingSoundtracks} and are marked with $\triangle$ to indicate so. We also provide oracle results based on the mixture, the ideal ratio mask, and the phase-sensitive filter~\cite{Erdogan2015Phase-sensitiveNetworks}. 

\subsection{Reducing Time-Frequency Modeling Complexity}
The first modification made to the original BSRNN (BSRNN-LSTM12) was to reduce the complexity of the time-frequency modeling module. Switching from LSTM to GRU and cutting the stack size down from 12 pairs to 8 pairs (BSRNN-GRU8) showed nearly no changes to the performance on average. While the GRU-based model performed slightly worse for dialogue, it performed better with effects than the LSTM-based modules. This switch allowed us to significantly cut down the parameters by almost \SI{40}{\percent}, while also reducing the considerable memory footprint during backpropagation. For this experiment, we used the Vocals V7 band definition from the original paper, which was used for both the ``vocals'' and ``other'' stem in MUSDB18, hence making it the most appropriate multi-purpose band definition for this analysis.

\subsection{Common Encoder}

The next modification was to merge the encoder section, that is, all modules up to and including the TF modeling module, into a shared system for all stems. This further cut the parameters down by \SI{45}{\percent} from BSRNN-GRU8. Again, the performance of this common-encoder model (BandIt) is still very similar to either BSRNN system on average. More interestingly, the performance in the effects stem increased by about \SI{1}{\decibel} compared to BSRNN-LSTM12, but this is also accompanied by a drop of about \SI{1}{\decibel} in dialogue stem performance. This seems to indicate that there is a slight competition in dynamically allocating information from three stems into the shared embedding. Qualitatively, however, speech is known to be easier to detect and semantically segment than effects due to the former being less acoustically diverse and more bandlimited on average. As such, since the speech performance at around \SI{13}{\decibel} is closer to the oracle performance, we consider the improvement in the effects stem performance of higher importance.

\subsection{Loss Function}

The next experiment is concerned with choosing the most appropriate loss function for the system. We experimented with 4 loss functions: the L1 loss, the mean squared error (MSE) loss, the proposed L1SNR loss, and the 2-norm ablation (L2SNR) of L1SNR. All loss functions were applied in the time domain, the real part of the spectrogram, and the imaginary part of the spectrogram like in \eqref{eq:loss}. Note that the distance function used in L2SNR is practically identical to commonly used negative SNR loss.

Training on L1SNR loss achieved the highest performance, with at least \SI{0.7}{\decibel} higher performance compared to L1 and L2SNR losses across all stems; the latter two performed similarly across all stems. MSE loss performed worst as expected, given that it has the weakest sparsity-encouraging property across the four losses. The order of the performance corroborates with our analyses in Section \ref{sec:pm}.\ref{ssec:loss}, but more thorough experiments will be needed in a separate work to fully verify our hypothesis. 

\subsection{Band Definitions}

We look into the five proposed overlapping-band definitions. For each band, we experimented with 48-band and 64-band variants. The 48-band variant has a larger input bandwidth per band but fewer neurons provisioned per linear frequency. Overall, the 64-band version consistently outperformed the corresponding 48-band counterpart of the same band type. Mel, TriBark, and ERB models tend to perform similarly. The similarity in performance between the three band types is not too surprising, given the similarity in both their nonlinear frequency transforms and filterbanks (see also \Cref{fig:bands}). In a 64-band setting, all band types performed better than the ideal ratio mask in the dialogue stem. In both 48- and 64-band settings, the musical band performed the best. We hypothesize that this is due to its underlying musical scale containing significantly more nonlinear-frequency units in the lower linear-frequency region than the other three scales, thus more channel capacity was provisioned to the information-dense lower linear-frequency region. 

For the best model at 100 epochs (Music 64), we let the model continue to train until the validation loss no longer improves for 20 epochs. This was achieved at epoch 278, with a total training time of about \SI{4.3}{days}. Per-epoch improvements after the first 100 epochs were very small, but accumulated to about \SI{0.5}{\decibel} improvement across all stems after the additional 178 epochs. The performance of this model (BandIt+) is also shown in \Cref{tab:results}.

\begin{table}[t]
    \setlength{\tabcolsep}{3pt}
    \caption{SNR (dB) Performance on MUSDB18-HQ Test Set.}
    \label{tab:musdb}
    \small
    \centering
    \begin{tabularx}{\linewidth}{
        X
        S[table-format=-2.1]S[table-format=-2.1]
        S[table-format=-2.1]S[table-format=-2.1]
        S[table-format=-2.1]S[table-format=-2.1]
        S[table-format=-2.1]S[table-format=-2.1]
        S[table-format=-2.1]S[table-format=-2.1]
    }
    \toprule
    Model & \multicolumn{1}{r}{Vocals} 
        & \multicolumn{1}{r}{Drums} 
        & \multicolumn{1}{r}{Bass}  
        & \multicolumn{1}{r}{Other}   
        & \multicolumn{1}{r}{Average}\\
    \midrule
    BandIt (Music 64, frozen enc.) &
    5.5 & \b 6.4 & \b 4.4 & \b 3.6 & \b 5.0\\
    Open-Unmix (\texttt{umxhq})  & 
    \b 6.0 & 5.6 & \b 4.4 & 3.4 & 4.9\\
    \bottomrule
    \end{tabularx}
\end{table}

\subsection{Generalizability}

We additionally tested the generalizability of the feature map learned by the encoder. This is done by freezing the encoder from the BandIt model with 64 musical bands and attaching a new randomly initialized decoder for an output stem that was not directly learned in the original 3-stem training. We first tested the generalizability on an ``easier'' task of obtaining the music-and-effects stem. Using the sum of the original music and effects stems outputs, the SNR and SI-SNR are at \SI{13.9}{\decibel} and \SI{13.7}{\decibel}, respectively. Training a new decoder for the composite stem achieves a slightly better output at \SI{14.1}{\decibel} for SNR and \SI{13.9}{\decibel} for SI-SNR.

Next, we trained new decoders on completely unseen music data from MUSDB18-HQ~\cite{musdb18}\footnote{The use of MUSDB18 here is strictly for the demonstration of model generalizability, and will not be used commercially.}. Note that MUSDB18 provides stereo data and the encoder was only trained on mono signals, so each channel of the music data was passed through the encoder independently. Despite only being trained to separate music as a whole without caring about its constituent instrumentals, the representations from the frozen encoder were sufficient to train decoders that are on par in performance to Open-Unmix, as shown in \Cref{tab:musdb}.

\subsection{Computational Complexity}
While the BandIt models have achieved state-of-the-art performance with lower overall complexity than BSRNN, it is important to note that the inference-time Flops count of a 64-band BandIt remains significantly higher than Hybrid Demucs, despite the latter having higher parameter counts, partially due to the RNN-heavy backbone of BandIt. Using 6-second chunk inputs on a machine with an Intel Core i9-11900K CPU and an NVIDIA GeForce RTX 3090 GPU, Demucs processed about 17.0 chunks per second on GPU while BandIt did so at about 8.7 chunks per second. On CPU, Demucs did so at about 1.1 chunks per second, while BandIt did so at about 0.3 chunks per second. The peak memory usage of BandIt at about 650 MB is slightly higher that than of Demucs at about 550 MB.

\section{Conclusion}

In this work, we propose BandIt, a generalization of the Bandsplit RNN to any complete or overcomplete partitions of the frequency axis. By also introducing a shared-encoder, a 1-norm SNR-like loss function, and psychoacoustically motivated band definitions, BandIt achieves state-of-the-art performance in CASS with fewer parameters than the original BSRNN or Hybrid Demucs. Future work includes more in-depth analysis of the behavior of the proposed loss function, deriving more information-theoretically optimal band definitions, and extending the work to more realistic audio data with more emotional, linguistic, and spatial diversity.

\section*{ACKNOWLEDGMENT}
The authors would like to thank 
Jordan Gilman,
Kyle Swanson, 
Mark Vulfson, and
Pablo Delgado for their assistance.

\FloatBarrier

\renewcommand*{\bibfont}{\small\setstretch{0.75}}
\bibliographystyle{IEEEbiba}
\bibliography{references}

\end{document}